# Speaker Verification in Emotional Talking Environments based on Three-Stage Framework

Ismail Shahin
Department of Electrical and Computer Engineering
University of Sharjah
Sharjah, United Arab Emirates
E-mail: ismail@sharjah.ac.ae

*Abstract*—This work is dedicated to introducing, executing, and assessing a three-stage speaker verification framework to enhance the degraded speaker verification performance in emotional talking environments. Our framework is comprised of three cascaded stages: gender identification stage followed by an emotion identification stage followed by a speaker verification stage. The proposed framework has been assessed on two distinct and independent emotional speech datasets: our collected dataset and Emotional Prosody Speech and Transcripts dataset. Our results demonstrate that speaker verification based on both gender cues and emotion cues is superior to each of speaker verification based on gender cues only, emotion cues only, and neither gender cues nor emotion cues. The achieved average speaker verification performance based on the suggested methodology is very similar to that attained in subjective assessment by human listeners.

## I. Introduction

Speaker verification is defined as the practice whether to accept or reject the requested speaker. It is considered as a true-or-false binary decision problem. Speaker verification technology appears in a wide range of applications such as: biometric person authentication, speaker verification for surveillance, forensic speaker recognition, and security applications including credit card transactions, computer access control, monitoring people, telephone voice authentication for long distance calling or banking access [1].

In terms of the spoken text, speaker verification comes in two forms: text-dependent and text-independent. In text-dependent, the same text is uttered in both training and testing phases, while in text-independent, there is no constraint of voice sample in the training and testing phases.

In this work, we address the problem of enhancing speaker verification performance in emotional environments based on proposing, applying, and testing a three-stage speaker verification framework which is made up of three sequential stages: gender identification stage followed by an emotion identification stage followed by a speaker verification stage.

## II. Prior Work

In speaker recognition community, there are large number of studies [2-6] that spot the light on speaker verification in emotional environments. The authors of [2] presented studies into the effectiveness of the state-of-the-art speaker verification techniques: Gaussian Mixture Model-Universal Background Model and Gaussian Mixture Model-Support Vector Machine (GMM-UBM and GMM-SVM) in mismatched noise conditions. The authors of [3] tested whether speaker verification algorithms that are trained in emotional environments give better performance when implemented to speech samples achieved under stressful or emotional conditions than those trained in a neutral environment only. Their conclusion is that training of speaker verification algorithms on a broader span of speech samples, including stressful and emotional conditions, rather than the neutral talking condition, is an encouraging method to improve speaker authentication performance [3]. The author of [4] proposed, applied, and evaluated a two-stage approach for speaker verification in emotional environments using completely Hidden Markov Models (HMMs). He examined the proposed approach using a collected speech dataset and obtained 84.1% as a speaker verification performance. The authors of [5] investigated the impact of emotion on the performance of an GMM-UBM based speaker verification system in such talking environments. In their work, they introduced an emotion-dependent score normalization method for speaker verification on emotional speech. They reported an average speaker verification performance of 88.5% [5]. In [6], the author focused on employing and evaluating a two-stage method to authenticate the claimed speaker in emotional environments. His method was made up of two recognizers which were combined and integrated into one recognizer using both HMMs and Suprasegmental Hidden Markov Models (SPHMMs) as classifiers. The two recognizers are: emotion identification recognizer followed by speaker verification recognizer. He attained average Equal Error Rate (EER) of 7.75% and 8.17% using a collected dataset and Emotional Prosody Speech and Transcripts (EPST) dataset, respectively.

Our current work mainly contributes to additional enhancement of speaker verification performance compared to that based on the two-stage methodology [6] by employing and evaluating a three-stage speaker verification framework to authenticate the claimed speaker in emotional environments. Our framework is comprised of three sequential recognizers that are combined and integrated into one recognizer using HMMs as classifiers in each stage. The three recognizers are: gender identifier followed by an emotion identifier followed by a speaker verifier. Specifically, our present work focuses on the enhancement of text-independent, gender-dependent, and emotion-dependent speaker verification performance in emotional environments.

The remaining of this article is arranged as follows: Section III explains the two speech datasets used to test the introduced framework and the extraction of features. The three-stage framework and the experiments are discussed in Section IV. The achieved results in the present study and their discussion are given in Section V. Finally, Section VI gives the concluding remarks of this work.

## III. SPEECH DATASETS AND EXTRACTION OF FEATURES

In this work, our proposed three-stage speaker verification method has been evaluated on two diverse and independent emotional speech datasets: our gathered dataset and Emotional Prosody Speech and Transcripts (EPST) dataset.

### A. Our Collected Dataset

Forty (twenty per gender) inexperienced mature (with ages ranging between 18 years and 55 years) native speakers of American English uttered the collected speech dataset in this work. The speakers were selected to naturally utter eight sentences and to keep away from overstressed expressions. Every speaker was requested to utter eight utterances where each utterance was spoken nine times under each of neutral, anger, sadness, happiness, disgust, and fear emotions. The eight utterances were chosen to be unbiased towards any emotion. These utterances are:
1) *He works five days a week.*
2) *The sun is shining.*
3) *The weather is fair.*
4) *The students study hard.*
5) *Assistant professors are looking for promotion.*
6) *University of Sharjah.*
7) *Electrical and Computer Engineering Department.*
8) *He has two sons and two daughters.*

The first four utterances of this dataset were utilized in the training session, whereas the remaining utterances were utilized in the evaluation session (text-independent problem). The collected dataset was recorded in an uncontaminated environment by a speech acquisition board using a 16-bit linear coding A/D converter and sampled at a sampling rate of 16 kHz. This dataset is a wideband 16-bit per sample linear data. A pre-emphasizer was applied to the speech signal samples. Afterwards, these signals were sliced into frames of 16 ms each where succeeding frames overlapped by 9 ms.

### B. Emotional Prosody Speech and Transcripts (EPST) Dataset

EPST dataset was introduced by Linguistic Data Consortium (LDC) [7]. This dataset was uttered by eight professional speakers (three male and five female) talking a sequence of semantically neutral utterances made up of dates and numbers spoken in fifteen distinct emotions including the neutral condition. Only six emotions (neutral, happiness, sadness, disgust, panic, and hot anger) were utilized in this study. In this dataset, only four utterances were utilized in the training session, while another different four utterances were utilized in the testing session.

### C. Extraction of Features

Mel-Frequency Cepstral Coefficients (MFCCs) have been utilized as the extracted features that characterize the phonetic content of the captured utterances in the two datasets. These features have been mainly utilized in speaker recognition [6], [8], [9], [10], [11] and emotion recognition [12], [13], [14], [15] studies. In this research, the number of states of HMMs is six.

## IV. THREE-STAGE SPEAKER VERIFICATION FRAMEWORK AND THE EXPERIMENTS

Our proposed framework assumes that *n* speakers are given for each gender where every talker emotionally talks in *m* emotions. Our overall suggested framework is composed of three cascaded and sequential stages as given in Fig. 1. The three stages are:

*First Stage: Gender Identification*

The first step of the entire three-stage method is to recognize the claimed speaker gender so the output of this phase becomes gender-dependent. Typically, automatic gender classification phase yields great performance without much work because the result of this phase is the claimed speaker either a male or a female. Thus, gender recognition problem is a binary categorization which is mostly not a very challenging step.

Two probabilities for each utterance are calculated in this stage using HMMs and the largest probability is selected as the recognized gender as shown in the coming equation,

$$G^* = \arg \max_{2 \geq g \geq 1} \left\{ P\left(O \middle| \Gamma^g\right) \right\} \quad (1)$$

where $G^*$ is the pointer of the recognized gender (male or female), $\Gamma^g$ is the $g^{th}$ HMM gender model, and $P(O|\Gamma^g)$ is the probability of the observation sequence $O$ that corresponds to the unidentified gender of the claimed speaker given the $g^{th}$ HMM gender model.

In the training phase of this step, the twenty male speakers producing entirely the first four utterances under the whole emotions of our dataset build the HMM male gender model, while the twenty female speakers producing completely the first four sentences under the entire emotions of our dataset derive the HMM female gender model. The overall number of utterances utilized to build each HMM gender model is 4320 (20 speakers × 4 sentences × 9 utterances/sentence × 6 emotions).

*Second Stage: Emotion Identification*

The aim of the present step is to recognize the undetermined emotion which corresponds to the claimed speaker who is talking emotionally provided his/her gender was recognized in the preceding stage. This step is termed gender-specific emotion identification. In the current step, there are *m* probabilities for each gender that are calculated using HMMs. The highest probability is selected as the recognized emotion for each gender as shown in the next equation,

$$E^* = \arg \max_{m \geq e \geq 1} \left\{ P\left(O \middle| G^*, \lambda_E^e\right) \right\} \quad (2)$$

where $E^*$ is the indicator of the recognized emotion, $\left(\lambda_E^e\right)$ is the $e^{th}$ HMM emotion model, and $P(O|G^*, \lambda_E^e)$ is the probability of the observation sequence $O$ that corresponds to the unspecified emotion provided the recognized gender and the $e^{th}$ HMM emotion model.

In the emotion identification stage, the $e^{th}$ HMM emotion model $\left(\lambda_E^e\right)$ for each gender has been constructed in the training session for each emotion utilizing the twenty speakers for each gender generating the entire first four utterances with a replication of nine utterances/sentence. The overall number of utterances utilized to construct every HMM emotion model for each gender is 720 (20 speakers × 4 sentences × 9 utterances/sentence).

*Third Stage: Speaker Verification*

The final stage of the overall suggested three-stage framework is to authenticate the speaker identity using HMMs provided that both of his/her gender and emotion were identified in the prior two stages (gender-specific and emotion-specific speaker verification problem) as presented in the following equation,

$$\Lambda(O) = \log\left[P\left(O \mid E^*, G^*\right)\right] - \log\left[P\left(O \mid \overline{E}^*, G^*\right)\right]$$
$$- \log\left[P\left(O \mid \overline{E}^*, \overline{G}^*\right)\right] \quad (3)$$

where $\Lambda(O)$ is the log-likelihood ratio in the *log* domain, $P(O|E^*,G^*)$ is the probability of the observation sequence $O$ that corresponds to the claimed speaker provided the true recognized emotion and the true recognized gender are given, $P(O|\overline{E}^*,G^*)$ is the probability of the observation sequence $O$ that corresponds to the claimed speaker given the false recognized emotion and the true recognized gender, and $P(O|\overline{E}^*,\overline{G}^*)$ is the probability of the observation sequence $O$ that corresponds to the claimed speaker provided the false recognized emotion and the false recognized gender. Equation (3) shows that the likelihood ratio is computed among model trained using data from recognized gender, recognized emotion, and claimed speaker.

The probability of the observation sequence $O$ which corresponds to the claimed speaker provided the correct recognized emotion and the true recognized gender can be obtained as [16],

$$\log P(O \mid E^*, G^*) = \frac{1}{T}\sum_{t=1}^{T} \log P(o_t \mid E^*, G^*) \quad (4)$$

where, $O = o_1 o_2 \ldots o_t \ldots o_T$.

The probability of the observation sequence $O$ which corresponds to the claimed speaker given the wrong recognized emotion and the true recognized gender can be obtained using a set of $B$ imposter emotion models: $\{\overline{E}_1^*, \overline{E}_2^*, \ldots, \overline{E}_B^*\}$ as,

$$\log P(O \mid \overline{E}^*, G^*) = \left\{\frac{1}{B}\sum_{b=1}^{B} \log\left[P(O \mid \overline{E}_b^*, G^*)\right]\right\} \quad (5)$$

where $P(O|\overline{E}_b^*,G^*)$ can be calculated using Equation (4). In our work, the value of $B$ is equal to $6 - 1 = 5$ emotions.

The probability of the observation sequence $O$ which corresponds to the claimed speaker provided the wrong recognized emotion and the false recognized gender can be determined utilizing the same set of $B$ imposter emotion models as,

$$\log P(O \mid \overline{E}^*, \overline{G}^*) = \left\{\frac{1}{B}\sum_{b=1}^{B} \log\left[P(O \mid \overline{E}_b^*, \overline{G}^*)\right]\right\} \quad (6)$$

where $P(O|\overline{E}_b^*,\overline{G}^*)$ can be calculated using Equation (4).

In the testing phase, every speaker of our dataset used nine utterances for every sentence of the last four sentences (text-independent) under each emotion. The overall number of utterances utilized in this phase is 8640 (40 speakers × 4 sentences × 9 utterances / sentence × 6 emotions). In this work, seventeen speakers per gender have been used as claimants and the remaining have been used as imposters.

## V. RESULTS AND DISCUSSION

In our work, a three-stage framework has been introduced, executed, and tested to increase the reduced speaker verification performance in emotional environments. Our introduced architecture has been evaluated on each of our collected and EPST datasets using HMMs as classifiers in each stage.

In this work, stage 1 of the whole proposed architecture gives 97.18% and 96.23% gender identification performance using the captured and EPST datasets, respectively. These two attained performances are larger than those obtained in some prior work [17], [18]. The authors of [17] obtained 92.00% as a gender identification performance in neutral talking environments. The authors of [18] achieved 90.26% as a gender identification performance using Berlin German dataset.

The second stage which is named gender-dependent emotion identification stage yields gender-dependent emotion identification performance based on HMMs and using each of the collected and EPST datasets as illustrated in Table 1. Based on this table, average emotion identification performance using the collected and EPST datasets is 83.03% and 83.08%, respectively. These two values are larger than those reported by the authors of [19] who reported a male and a female average emotion identification performance of 61.10% and 57.10%, respectively.

Table 1
Gender-dependent emotion identification performance using each of the captured and EPST datasets

|  | Emotion identification performance (%) | |
| --- | --- | --- |
| Emotion | Collected dataset | EPST dataset |
| Neutral | 92.2 | 92.3 |
| Anger | 78.2 | 77.9 |
| Sadness | 81.9 | 81.1 |
| Happiness | 85.4 | 84.9 |
| Disgust | 79.2 | 79.6 |
| Fear | 81.3 | 82.7 |

Table 2 yields percentage Equal Error Rate (EER) of speaker verification in emotional environments based on the overall three-stage framework using each of the captured and EPST datasets. The average percentage EER is 9.50% and 10.00% using the collected and EPST datasets, respectively. These averages are less than those reported based on the two-stage framework proposed by the author of [6]. This table shows that the least percentage EER takes place when speakers talk neutrally, whereas the largest percentage EER happens when speakers talk angrily. This table evidently yields higher percentage EER when speakers speak emotionally compared to when speakers speak neutrally. This is because the presented percentage EER in Table 2 is the resultant of percentage EER of each stage of the three-stage method. The three-stage framework could have a destructive effect on the overall speaker verification performance particularly when both the gender (stage 1) and emotion (stage 2) of the claimed speaker has been falsely recognized.

Table 2
Percentage EER based on the three-stage framework using the captured and EPST datasets

|  | EER (%) | |
| --- | --- | --- |
| Emotion | Collected dataset | EPST dataset |
| Neutral | 3.0 | 3.5 |
| Anger/Hot Anger | 11.5 | 12.5 |
| Sadness | 10.5 | 11.5 |
| Happiness | 10.5 | 11.0 |
| Disgust | 11.0 | 12.0 |
| Fear/Panic | 10.5 | 9.5 |

In the current work, the achieved average percentage EER based on the three-stage architecture is less than that attained in prior studies:

1) The author of [4] obtained 15.9% as an average percentage EER in emotional environments using HMMs only.
2) The authors of [10] reported an average percentage EER of 11.48% in emotional environments using GMM-UBM based on emotion-independent method.

Five major experiments have been done in the present study to test the achieved results based on the three-stage architecture. The five experiments are:

(1) Experiment 1: The percentage EER based on the proposed three-stage architecture has been compared with that based on the one-stage framework (text-independent, gender-independent, and emotion-independent speaker verification) using independently each of the captured and EPST datasets. Based on the one-stage method and using HMMs as classifiers, the average percentage EER is 14.75% and 14.58% using the captured and EPST datasets, respectively. Therefore, we can conclude based on this experiment that the three-stage speaker verification architecture is superior to the one-stage speaker verification framework. Hence, embedding both of gender and emotion recognition steps into the one-stage speaker verification architecture in emotional environments significantly improves speaker verification performance competed to that without embedding these two stages.

(2) Experiment 2: The percentage EER using the proposed three-stage framework has been competed with that based on the emotion-independent two-stage framework (text-independent, gender-dependent, and emotion-independent speaker verification) using independently each of the captured and EPST datasets. Based on this framework, the average percentage EER based on the text-independent, gender-dependent, and emotion-independent method is 13.09% and 12.98% using, respectively, the collected and EPST datasets. Therefore, inserting emotion identification stage into the emotion-independent two-stage speaker verification architecture in emotional environments considerably enhances speaker verification performance competed to that without such a stage. Hence, adding emotion identification stage into the one-stage speaker verification architecture in emotional environments noticeably increases speaker verification performance competed to that without adding this stage.

(3) Experiment 3: The percentage EER using the introduced three-stage framework has been compared with that based on the gender-independent two-stage framework (text-independent, gender-independent, and emotion-dependent speaker verification) using individually each of the captured and EPST datasets. Based on this methodology, the average percentage EER is 12.05% and 11.88% using the collected and EPST datasets, respectively. Consequently, adding gender identification stage into the gender-independent two-stage speaker verification architecture in emotional environments appreciably improves speaker verification performance competed to that without adding this stage.

(4) Experiment 4: The overall three-stage architecture has been tested for the worst-case scenario. This scenario takes place when stage 3 gets incorrect input from both the preceded two stages (stage 1 and stage 2). Hence, this scenario happens when speaker verification stage receives false identified gender and wrong recognized emotion. The attained average percentage EER in the worst-case scenario based on HMMs is 15.12% and 15.02% using the captured and EPST datasets, respectively.

These attained averages are very similar to those obtained using the one-stage approach (14.75% and 14.58% using the captured and EPST datasets, respectively).

(5) Experiment 5: An informal subjective assessment of the suggested three-stage framework has been implemented with five male and five female nonprofessional listeners using the collected speech dataset. These listeners were arbitrarily chosen from distinct ages ($20 - 50$ years old). These judges were not used in collecting the collected dataset. An overall of 960 utterances (20 speakers $\times$ 2 genders $\times$ 6 emotions $\times$ the last 4 sentences of the dataset) have been utilized in this experiment. Each listener in this assessment is asked three sequential questions for each test sentence. The three consecutive questions are: recognize the unidentified gender of the claimed speaker, afterwards, recognize the unknown emotion of the claimed speaker given his/her gender was recognized, and finally verify the claimed speaker provided both his/her gender and emotion were identified. Based on the subjective evaluation of this experiment, the average: gender identification performance, emotion identification performance, and speaker verification performance is 96.24%, 87.57%, and 84.37%, respectively. These averages are very alike to those attained based on the novel three-stage speaker verification architecture.

VI. CONCLUDING REMARKS

In this study, a novel three-stage speaker verification framework has been introduced, implemented, and assessed to increase the low speaker verification performance in emotional environments. This architecture combines and integrates three cascaded recognizers: gender identifier, followed by emotion identifier, followed by speaker verifier into one recognizer using HMMs as classifiers in every stage. This architecture has been assessed on two distinct and independent speech datasets: the captured and EPST. Five major experiments have been done in the current study to test the proposed framework. Some concluding remarks can be obtained in our research. Firstly, speaker verification in emotional environments based on both gender cues and emotion cues leads each of that based on gender cues only, emotion cues only, and neither gender cues nor emotion cues. Secondly, the three-stage framework works nearly the same as the one-stage method when the third stage of the three-stage architecture receives both an incorrect recognized gender and an incorrect recognized emotion from the preceded two stages. Thirdly, emotion cues are more important than gender cues to speaker verification system. However, both of gender and emotion cues are more prominent than emotion cues only to speaker verification system in these talking environments. Finally, this study apparently demonstrates that the emotional status of the claimed speaker has a negative impact on speaker verification performance.

Our proposed three-stage speaker verification method has some limitations. First, in the three-stage architecture, the needed processing calculations and the time spent are higher than those in the one-stage framework. Second, speaker verification performance using the three-stage architecture is imperfect. This three-stage performance is the resultant of three non-ideal performances:

(a) The unknown gender of the claimed speaker is not 100% correctly identified in the first stage.
(b) The unknown emotion of the claimed speaker is imperfectly recognized in stage 2.
(c) The claimed speaker is non-ideally verified in the last stage.

For future work, our plan is to additionally alleviate speaker verification performance degradation in emotional environments by proposing novel classifiers. Our plan also is to analytically work on the three-stage architecture to determine the performance of each stage individually and the overall performance of the three-stage speaker verification architecture; we intend to develop a mathematical relationship between the whole performance and each stage performance.

ACKNOWLEDGMENT

The authors wish to thank University of Sharjah for funding this work through the competitive research project entitled "Emotion Recognition in each of Stressful and Emotional Talking Environments Using Artificial Models", No. 1602040348-P.

REFERENCES

[1] D. A. Reynolds, "An overview of automatic speaker recognition technology," ICASSP 2002, Vol. 4, May 2002, pp. IV-4072- IV-4075.

[2] S. G. Pillay, A. Ariyaeeinia, M. Pawlewski, and P. Sivakumaran, "Speaker verification under mismatched data conditions," IET Signal Processing, Vol. 3, issue 4, July 2009, pp. 236-246.

[3] K.R. Scherer, T. Johnstone, G. Klasmeyer, and T. Banziger, "Can automatic speaker verification be improved by training the algorithms on emotional speech?," Proceedings of International Conference on Spoken Language Processing, October 2000, Vol. 2, pp. 807-810.

[4] I. Shahin, "Verifying speakers in emotional environments," The 9th IEEE International Symposium on Signal Processing and Information Technology, Ajman, United Arab Emirates, December 2009, pp. 328-333.

[5] W. Wu, T. F. Zheng, M. X. Xu, and H. J. Bao, "Study on speaker verification on emotional speech," INTERSPEECH 2006 – Proceedings of International Conference on Spoken Language Processing, September 2006, pp. 2102-2105.

[6] I. Shahin, "Employing emotion cues to verify speakers in emotional talking environments," Journal of Intelligent Systems, Special Issue on Intelligent Healthcare Systems, DOI: 10.1515/jisys-2014-0118, Vol. 25, issue 1, January 2016, pp. 3-17.

[7] Emotional Prosody Speech and Transcripts dataset. www.ldc.upenn.edu/Catalog/CatalogEntry.jsp?catalogId=LDC2002S28. Accessed 20 April, 2017.

[8] I. Shahin, "Identifying speakers using their emotion cues," International Journal of Speech Technology, Vol. 14, No. 2, June 2011, pp. 89 – 98, DOI: 10.1007/s10772-011-9089-1.

[9] I. Shahin, "Employing both gender and emotion cues to enhance speaker identification performance in emotional talking environments," International Journal of Speech Technology, Vol. 16, issue 3, September 2013, pp. 341-351, DOI: 10.1007/s10772-013-9188-2.

[10] W. Wu, T. F. Zheng, M. X. Xu, and H. J. Bao, "Study on speaker verification on emotional speech," INTERSPEECH 2006 – Proceedings of International Conference on Spoken Language Processing (ICSLP), September 2006, pp. 2102-2105.

[11] T. H. Falk and W. Y. Chan, "Modulation spectral features for robust far-field speaker identification," IEEE Transactions on Audio, Speech and Language Processing, Vol. 18, No. 1, January 2010, pp. 90-100.

[12] I. Shahin, "Speaker identification in emotional talking environments based on CSPHMM2s," Engineering Applications of Artificial Intelligence, Vol. 26, issue 7, August 2013, pp. 1652-1659, http://dx.doi.org/ 10.1016/ j.engappai. 2013.03.013.

[13] C. M. Lee and S. S. Narayanan, "Towards detecting emotions in spoken dialogs," IEEE Transactions on Speech and Audio Processing, Vol. 13, No. 2, March 2005, pp. 293-303.

[14] N. Sato and Y. Obuchi, "Emotion recognition using Mel-frequency cepstral coefficients," Journal of Natural Language Processing, Vol. 14, No. 4, 2007, pp. 83-96.

[15] I. Shahin, "Gender-dependent emotion recognition based on HMMs and SPHMMs," International Journal of Speech Technology, Vol. 16, issue 2, June 2013, pp. 133-141, DOI: 10.1007/s10772-012-9170-4.

[16] D. A. Reynolds, "Automatic speaker recognition using Gaussian mixture speaker models," The Lincoln Laboratory Journal, Vol. 8, No. 2, 1995, pp. 173-192.

[17] H. Harb and L. Chen, "Gender identification using a general audio classifier," International Conference on Multimedia and Expo 2003 (ICME '03), July 2003, pp. II – (733 – 736).

[18] T. Vogt and E. Andre, "Improving automatic emotion recognition from speech via gender differentiation," Proceedings of Language Resources and Evaluation Conference (LREC 2006), Genoa, Italy, 2006.

[19] D. Ververidis and C. Kotropoulos, "Emotional speech recognition: resources, features, and methods," Speech Communication, Vol. 48, issue 9, September 2006, pp. 1162-1181.

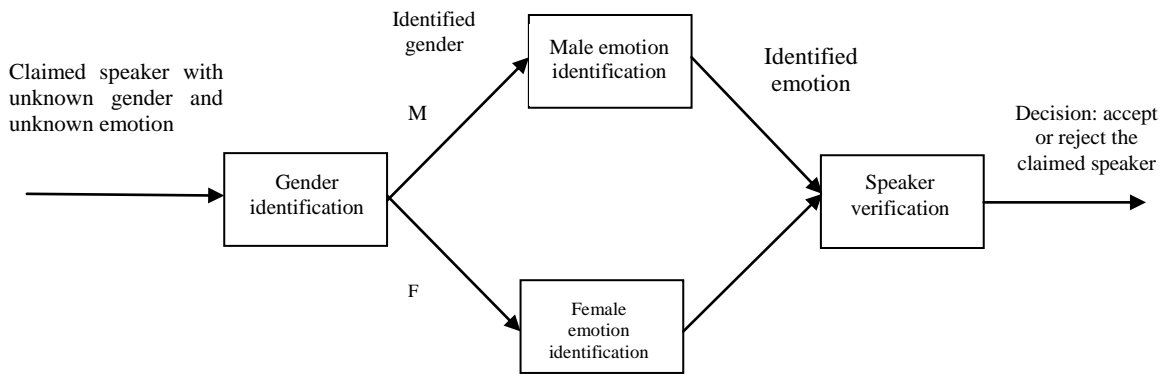

Fig. 1. Block diagram of the overall proposed three-stage speaker verification framework